# Mobile IoT device for monitoring people with bpm problems.

Luis Chuquimarca
*Universidad Estatal Península de Santa Elena*
*Facultad de Sistemas y Telecomunicaciones*
Santa Elena, Ecuador
lchuquimarca@upse.edu.ec

Dahyana Roca
*Universidad Estatal Península de Santa Elena*
*Facultad de Sistemas y Telecomunicaciones*
Santa Elena, Ecuador
dahyanarocasoriano@hotmail.com

Washington Torres
*Universidad Estatal Península de Santa Elena*
*Facultad de Sistemas y Telecomunicaciones*
Santa Elena, Ecuador
wtorresguin@gmail.com

Luis Amaya
*Universidad Estatal Península de Santa Elena*
*Facultad de Sistemas y Telecomunicaciones*
Santa Elena, Ecuador
lamaya@upse.edu.ec

Jaime Orozco
*Universidad Estatal Península de Santa Elena*
*Facultad de Sistemas y Telecomunicaciones*
Santa Elena, Ecuador
jorozco@upse.edu.ec

Davis Sánchez
*Universidad Estatal Península de Santa Elena*
*Facultad de Sistemas y Telecomunicaciones*
Santa Elena, Ecuador
dsanchez@upse.edu.ec

**Abstract—** The developed system using a mobile electronic device for monitoring and warnings of heart problems, when the heart rate is outside the nominal range, which ranges from 60 to 100 beats per minute. Also, a system has been developed to save and monitor in real time changes of the cardiac pulsations, through a sensor connected to a control system. The connection of the communication module for Arduino GSM/GPRS/GPS, using the GPS network to locate the user. In addition, this device connects with GSM / GPRS technology that allows text messages to be sent to the contact number configured in the device, when warnings of heart problems are issued, moreover connects to the internet to store data in the cloud.

## I. INTRODUCTION

Remote monitoring as an application of telemedicine is widely considered as an important area in the future of modern medicine [1]. The cardiovascular diseases (CVD) are one of the leading causes of death worldwide [2]. Therefore, continuous monitoring of patients is a possible solution. Heart disease as heart arrhythmias, heart attack, coronary heart disease, congestive heart failure and congenital heart disease are some of the cause of death for men and women in the world [3] [4].

In this proposed device, the heartbeat is measured using a bpm sensor, which receives the analog signal then converts to digital signal, using the analog-to-digital converter (ADC), also has a filter and amplification system which allows the data to be obtained and processed by the microcontroller. In addition, when there are abnormalities in the heartbeat, it sends pop-up alerts via Short Message Service (SMS), through the GSM communication module and stores the data in an ThingSpeak web platform [5] [6].

The present paper is divided into three sections: the second section, describes the proposed system, the third section, the content development, fourth section, presents the methodology of project. Finally, the paper presents the conclusions concerning the development, the design and its possible benefits.

## II. PROPOSED SYSTEM

First establishes the connection for the data acquisition system using an ATmega32u4 microcontroller, the pulse cardiac sensor and screen OLED for indicate information of data emitted by the sensor, each time the touch button is pressed. Also, in the connection of the second ATmega32u4 microcontroller and the module GSM/GPRS/GPS which constitutes the communication system (see Fig. 1).

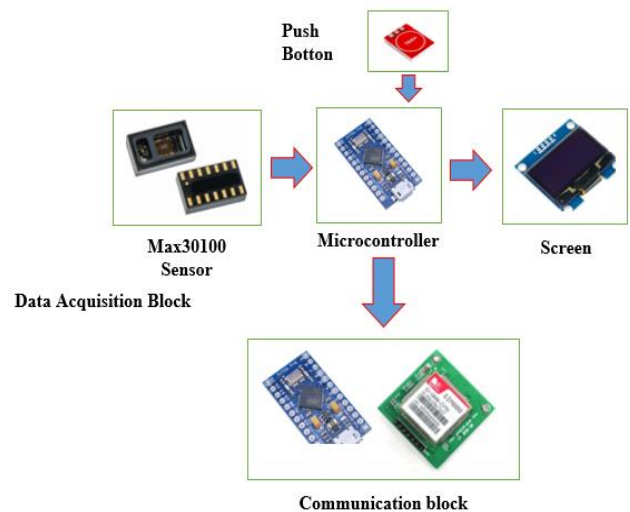

Fig. 1 Data acquisition system block diagram.

Data acquisition is constantly done through the MAX30100 integrated chip, which can reduce circuit design and system power consumption. The signal processing methods, such as amplification and filtering systems are performed using the libraries of the MAX30100 sensor [7], so that the sensor data can be obtained and processed by the ATmega32u4 microcontroller, which contains the programming algorithm necessary for the system, through i2c communication with the predetermined ports, in addition the data is sent to the microcontroller that contains the communication module and alert system (see Fig. 1).

For the connection between the data acquisition system and the communication system (see Fig. 2), considering the communication ports for the transmission / reception of predetermined data between the SIM808 module and the second ATmega32u4 microcontroller, using serial communication to send data collected by the sensor, they are the default transmit / receive ports on the microcontroller.

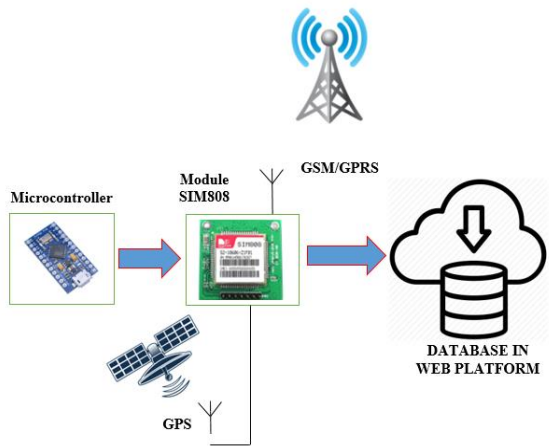

Fig. 2 Communication system block diagram.

When there is an alteration of the heart rate, the alert system is activated, which It would send an SMS with the information collected from the sensor and the location of the person obtained by the GPS module [8].

## III. CONTENT DEVELOPMENT

### A. Cardiac pulse oximeter

The MAX30100 sensor has two infrared LEDs with a wavelength of 660nm and 940nm, photodetector, optimized optics, and 16 bits ADC (see Fig. 3). The sensor operates at 5V power supplies and it turns off through software [9]. The MAX30100 provides solution by its very small size without decrease optical or electrical performance (see Fig. 3). It configure through libraries and digital output data is stored in a First-In, First-Out (FIFO) within the device, allowing the MAX30100 to connect to a microcontroller on a shared bus, where data is not read continuously from the device registers [7] [10].

To measure heart rate, the change in blood flow is detected in areas where the skin is not thick, for example, in the wrist where the sensor is in contact with the skin and can detect the blood flow, the data on heart rate is obtained together with the saturation of blood (SpO2) (see Fig. 3) [9].

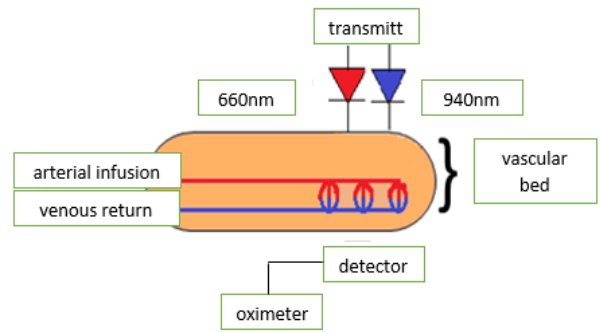

Fig. 3: Pulse oximeter sensor.

### B. Atmel Microcontroller.

The ATmega32u4 provides UART – TTL (5 V) communication, through serial zero (Tx) and one (Rx) pins. Through this configuration, the communication between the Arduino Pro Micro module and the pulse oximeter sensor is done, in addition a second Arduino Pro Micro module has used (see Fig. 4), which has the algorithm that allows communication with the SIM808 communication module [11].

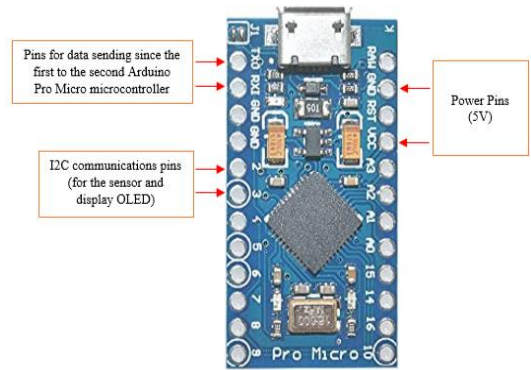

Fig. 4 Pins used on the Arduino Pro Micro.

### C. GSM/GPRS/ GPS Module.

GPRS quad band module that combines GPS technology for satellite navigation. The compact design integrates GPRS and GPS in a Surface-Mount Technology (SMT) package that significantly saves time and costs so that customers have developed applications that allow trouble-free tracking in different locations and at any time with signal coverage [12].

It is a module that allows access to the GSM/GPRS connection compatible with the four frequency bands for the GSM network (850, 900, 1800, 1900 MHz), it also combines GPS technology to obtain the latitude and length position. Its module incorporates a low energy consumption mode and connects with lithium battery [13]. The SIM808 module is configured using AT command and a serial communication interface. The SIM808 operate with a voltage of 5 V to 10 V (see Fig. 5). Once the module is powered and connects the GSM and GPS antennas and in addition to the SIM card correctly, the LED flashes slowly, which indicates that the module is registered in the network, allowing the user to make any of the configurations that are enabled this module [14].

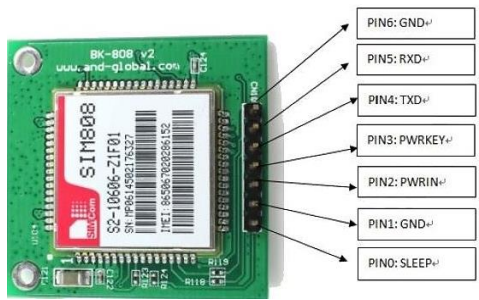

Fig. 5 GSM/GPRS/GPS module .

### D. IoT ThingSpeak platform.

The ThingSpeak platform web is used to store the bpm and SpO2 data emitted by the sensor. Moreover, it provides applications that allow real-time data visualization through a graphical interface (see Fig. 6) [15] [16].

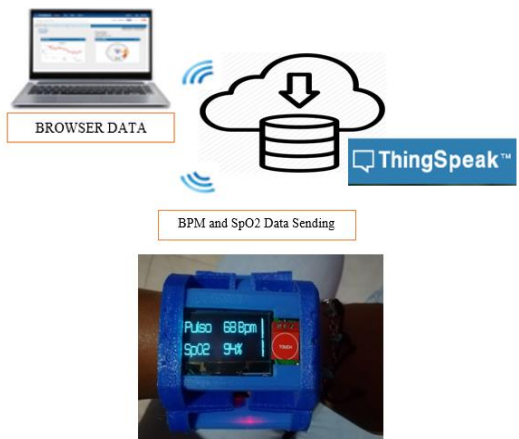

Fig. 6 Sending bpm and SpO2 data to the ThingSpeak platform

## IV. METHODOLOGY

For the development of the project considering each one of the requirements, an implementation process was carried out to guarantee its correct functionality [17].

### A. Physical design of the device

The development of the cardiac pulse detection device has a interconnected four-part design and the dimensions are 40 mm long, 52 mm wide and 12 mm thick (see Fig. 7).

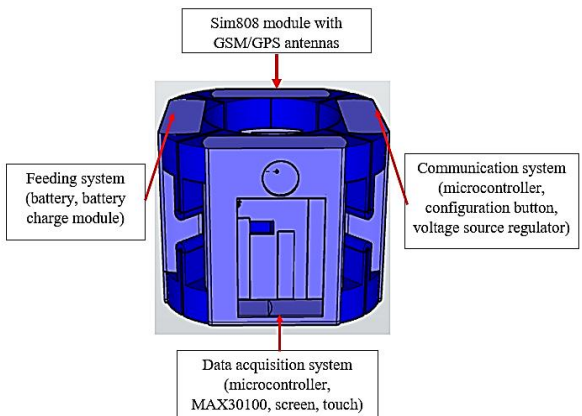

Fig. 7 Physical design of the device.

The physical structure of the device was designed for the location of each mentioned element using the SketUp software [18], and the parts of the device were printed in polylactic acid (PLA) material, so that the device has a comfortable aspect for the user (see Fig. 8).

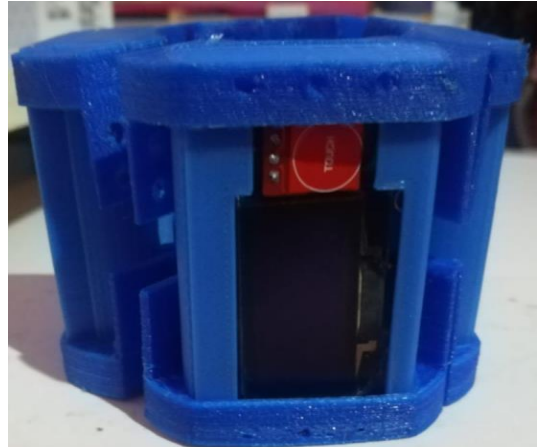

Fig. 8 Physical design of the device.

### B. Configuration device

The methodology of the implementation of device as follows:

The emergency alert is generated when the bpm value is out of nominal range (60 to 100 bpm) (see Fig. 9) [19].The parameters recognized by the microcontroller memory can be configured manually by the user. These parameters are phone numbers to whom the pop-up alert is issued, including the Apikey of the web platform, which is an id to register the data in the web platform (see Fig. 9). The configuration of the parameters is done pressing the main button for "configuration" mode, with a period of 80 seconds allowing the user to register the contact numbers, sending text messages to the sim number inserted in the SIM808 module.

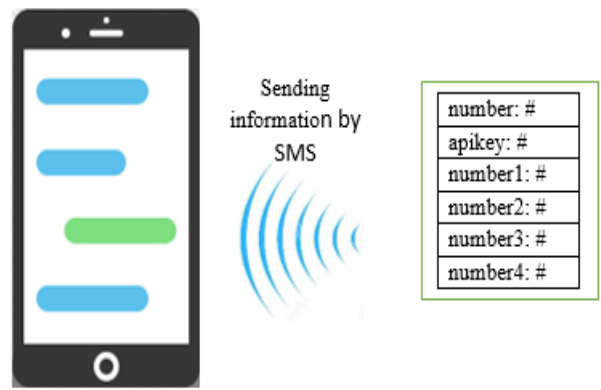

Fig. 9 Parameter configuration through text messages

## V. RESULTS

### A. Comparison between bpm measurement device.

The following results were obtained with the measurements of the developed device and a "Scian" device to determine the functionality of the device compared to market products (see Fig. 10) [20].

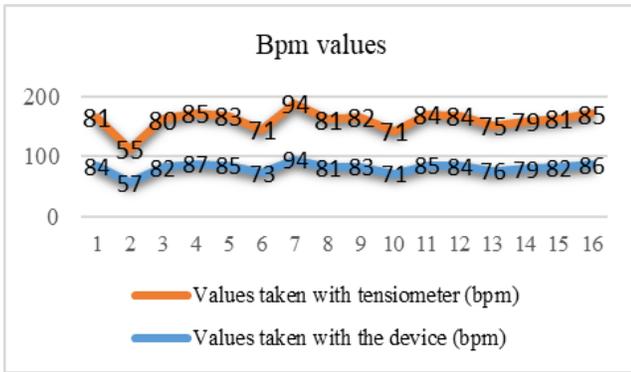

Fig. 10 Bpm values taken with the developed device compared to a "Scian" digital blood pressure monitor

*B. Energy consumption*

The development device operates with a LiPo battery that supplies a voltage of 3.7 - 4.2 V with full charge and an amperage of 1800 mA, when the device is powered, the current consumption of the device is 200 mA, whereas the sensor, displayed and communication module constantly operating, the calculations of the battery life is as follows.

$$t = \frac{1800 \text{ mAh}}{200 \text{ mA}} \approx 9 \text{ OPERATING HOURS}$$

According to this calculation the device has approximately nine hours of functionality.

*C. Alert Messages*

When the bpm value varies outside the nominal range, which are between 60 and 100 bpm, the alert message contains the value of bpm, SpO2 and a Uniform Resource Locator (URL) that contains the person's location, , through the Google Maps tool, in this way the user receives faster medical assistance (see Fig. 11).

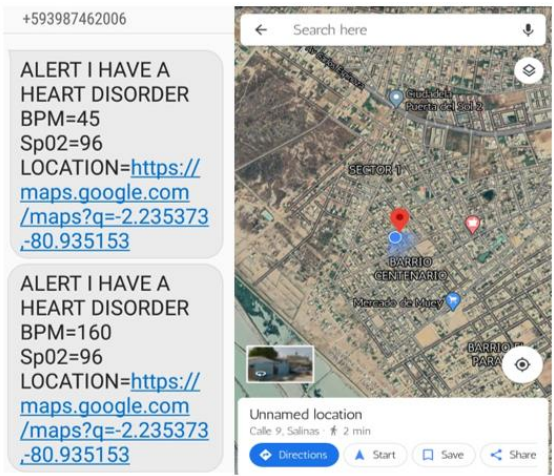

Fig. 11 Alert message with each of the parameters set.

*D. Sending data to the IoT platform*

The result can be verified through the following image that contains information of the data sent through the serial port on the Arduino platform, in addition to the information acquired in real time displayed on the ThingSpeak platform (see Fig. 12).

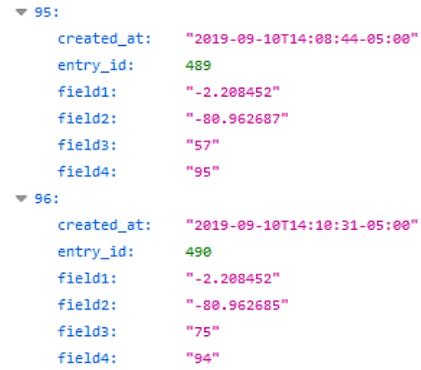

Fig. 12 Information visualization on the ThingSpeak platform.

The next illustration the information received on the ThingSpeak platform, which can be a modification by minutes, hours or days (see Fig. 13).

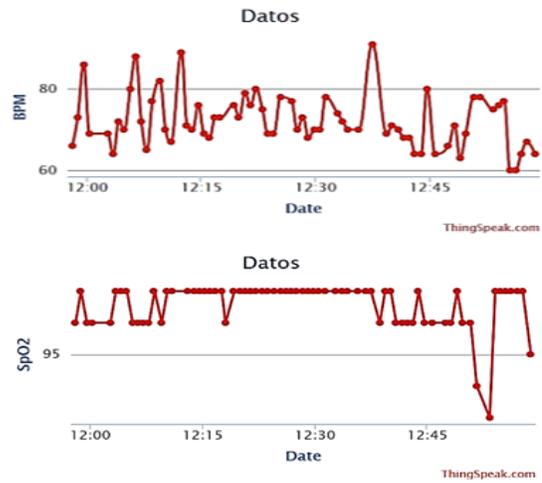

Fig. 13 Visualization of bpm and SpO2 data for minutes on the ThingSpeak platform.

The data to the web platform is sent with a time interval of 48 seconds. Therefore, the operating percentage of one hour is found, where the 75 shipments to the ThingSpeak platform, only 73 shipments was received.

The data to the web platform is sent with a time interval of 48 seconds. Therefore, the operating percentage of one hour is found, where the 75 shipments to the ThingSpeak platform, only 73 shipments was received.

*E. Mobile data consumption*

According to the Android application, the mobile data consumption was approximately 123.70 Kbytes in one hour. Therefore, there could be an approximate consumption of 0.1237 Mbytes per hour, and a value of 2.9688 Mbytes in 24 hours of device use.

VI. CONCLUSION

The heart rate detector device was developed using the MAX30100 sensor for its small size and high performance, configured with a wavelength of 660 nm for greater efficiency. In addition, the structure of the device was designed and implemented with PLA materials using a 3D printer, the dimensions are 40 mm long, 52 mm wide and 12 mm thick, and it is located on the person's wrist. A pop-up alert was programmed to detect bpm outside the range of 60 -100 bpm, so the person's location and bpm information is sent via SMS to the contact numbers configured in the device's internal memory. The 73 data packet deliveries were made to the ThingSpeak web platform every 48 seconds, using 0.12 Mbytes of mobile data in one hour of device operation.